\documentclass[useAMS,showkeys]{tMPH2e}

\usepackage{amsmath}

\newcommand\beq{\begin{equation}}
\newcommand\eeq{\end{equation}}
\newcommand\beqa{\begin{eqnarray}}
\newcommand\eeqa{\end{eqnarray}}
\newcommand{\nn}{\nonumber\\}

\newcommand{\mo}{\mu_1}
\newcommand{\mt}{\mu_2}
\newcommand{\mth}{\mu_3}
\newcommand{\mn}{\mu_n}
\newcommand{\pure}{{\text{p}}}

\begin{document}



\title{Test of a universality ansatz for the contact values
of the radial distribution functions of hard-sphere mixtures near a
hard wall}



\author{Mariano L\'{o}pez de Haro\thanks{$^\ast$
Corresponding author. Email: malopez@servidor.unam.mx}$^\ast$
 \\\vspace{6pt}
Centro de Investigaci\'on en Energ\'{\i}a, UNAM, Temixco, Morelos
62580, M{e}xico \\
\vspace{6pt} Santos B. Yuste\thanks{$^\dagger$ Email:
santos@unex.es}$^\dagger$ and Andr\'es Santos\thanks{$^\ddagger$
Email: andres@unex.es\vspace{10pt}
\newline\centerline{\tiny{ {\em Molecular Physics}}}
\newline\centerline{\tiny{ISSN 0026-8976 print/ ISSN
1362-3028
 online
\textcopyright 2005 Taylor \& Francis Ltd}}
\newline\centerline{\tiny{ http://www.tandf.co.uk/journals}}
\newline \centerline{\tiny{DOI:
10.1080/002689700xxxxxxxxxxxx}}}$^\ddagger$\\\vspace{6pt}
Departamento de F\'{\i}sica, Universidad de Extremadura, E-06071
Badajoz, Spain}

\received{\today}

\label{firstpage} \doi{10.1080/002689700xxxxxxxxxxx}

\issn{1362-3028}  \issnp{0026-8976} 

\markboth{L\'opez de Haro, Yuste and Santos}{Test of a universality
ansatz for the contact values of the radial distribution functions
of hard-sphere mixtures near a hard wall}

\maketitle
\begin{abstract}
Recent Monte Carlo simulation results for the contact values of
polydisperse hard-sphere mixtures at a hard planar wall are
considered in the light of a universality assumption made in
approximate theoretical approaches. It is found that the data seem
to fulfill the universality ansatz reasonably well, thus opening up
the possibility of inferring properties of complicated systems from
the study of simpler ones.
\end{abstract}

\section{Introduction}
\label{sec1}

In hard-sphere systems, the general statistical mechanical relation
between the thermodynamic properties and the structural properties
takes a rather simple form. Since the internal energy in these
system reduces to the one of the ideal gas and the pressure equation
only involves the contact values of the radial distribution
functions (rdf), a knowledge of such contact values is enough to
obtain their equation of state (EOS) and all their thermodynamic
properties. However, such a program can not be carried out
analytically due to the lack of exact expressions for these contact
values up to the present day. Under these circumstances, the best
one can do is to rely on sensible (approximate) proposals based on
as sound as possible theoretical results or to rely on computer
simulation values. Clearly, the situation is rather more complicated
for mixtures than for a single component fluid and in fact for this
latter many accurate (albeit empirical) equations of state have
appeared in the literature from which the contact value may be
readily derived.

A key analytical result is due to Lebowitz \cite{L64}, who obtained
the exact solution of the Percus--Yevick (PY) equation of additive
hard-sphere mixtures and provided explicit expressions for the
contact values of the rdf. Also analytical are the contact values of
the Scaled Particle Theory (SPT) \cite{LHP65,R88}. Neither the PY
nor the SPT lead to accurate values and so Boubl\'{\i}k \cite{B70}
(and, independently, Grundke and Henderson \cite{GH72} and Lee and
Levesque \cite{LL73}) proposed an interpolation between the PY and
the SPT contact values, that we will refer to as the BGHLL contact
values, which leads to the widely used and rather accurate
Boubl\'{\i}k--Mansoori--Carnahan--Starling--Leland (BMCSL) EOS
\cite{B70,MCSL71} for hard-sphere mixtures. Refinements of the BGHLL
values have been subsequently introduced, among others, by Henderson
\textit{et al.} \cite{Henderson}, Matyushov and Ladanyi  \cite{ML97}
and Barrio and Solana \cite{BS00} to eliminate some drawbacks of the
BMCSL EOS in the so-called colloidal limit of {binary} hard-sphere
mixtures. On a different path but also having to do with the
colloidal limit, Viduna and Smith \cite{VS02} have proposed a method
to obtain contact values of the rdf of hard-sphere mixtures from a
given EOS. In previous work we have made proposals for the contact
values of the rdf valid for mixtures with an arbitrary number of
components and in arbitrary dimensionality \cite{SYH99,Contact} and
for a hard-sphere polydisperse fluid \cite{SYH05}, that require as
the only input the EOS of the one-component fluid. Apart from
satisfying known consistency conditions,  they are sufficiently
general and flexible to accommodate any given EOS for the
one-component fluid. As far as computer simulation results are
concerned, contact values of the rdf of hard-sphere systems have
been reported by Lee and Levesque \cite{LL73}, Baro\v{s}ov\'a
\textit{et al.} \cite{BMLS96}, Lue and Woodcock \cite{LW99}, Cao
\textit{et al.} \cite{CCHW00}, Henderson \textit{et al.}
\cite{HTWC05}, Buzzacchi \textit{et al.} \cite{Ignacio} and
Malijevsk\'y \cite{Alexander}. In particular, these last two
references study hard-sphere mixtures in the presence of a hard
planar wall.

It is interesting to point out that in the case of multicomponent
mixtures of hard spheres and in the polydisperse hard-sphere fluid,
the contact values which follow from the solution of the PY equation
\cite{L64}, those of the SPT approximation \cite{LHP65,R88}, those
of the BGHLL interpolation \cite{B70,GH72,LL73} and our own
prescriptions \cite{SYH99,Contact,SYH05} exhibit a feature that one
might catalogue as a ``universal'' behavior because, once the
packing fraction is fixed, the expressions for the contact values of
the rdf for all pairs of like and unlike species depend on the
diameters of both species and on the size distribution \textit{only}
through a single dimensionless parameter, irrespective of the number
of components in the mixture.

In Ref.\ \cite{SYH05} we have presented a comparison of the
different theoretical proposals for the contact values of the rdf
and the ensuing EOS stemming from them with the simulation results.
The aim of the present paper is to assess, apart from the accuracy
of such proposals, whether the universality feature alluded to above
is indeed present in the simulation data in the case of mixtures in
the presence of a hard wall. This represents an extreme case and
therefore a proper test ground for our approach.

The paper is organized as follows. In order to make the paper
self-contained, in Sec.\ \ref{sec2} we rederive our most recent
proposal \cite{SYH05} for the contact values of the rdf (labelled as
e3 for the reasons explained below) using some known consistency
conditions and two different routes to compute the compressibility
factor of a polydisperse hard-sphere system in the presence of a
hard planar wall. We also point out in this section that the other
proposals sharing the universality feature, namely the  PY, SPT,
BGHLL and our two previous proposals \cite{SYH99,Contact} may be
cast in the same form as our e3 proposal, but that only the SPT and
the e3 proposals are consistent in the sense that they lead to the
same compressibility factors with the two different routes. Section
\ref{sec3} deals with the comparison between the various contact
values and simulation results, examining both the accuracy of the
theories as well as whether the universality ansatz is confirmed by
the simulation data. We close the paper in Sec.\ \ref{sec4} with
further discussion and some concluding remarks.

\section{Contact values of the radial distribution functions}
\label{sec2} Consider a polydisperse hard-sphere mixture with a
given size distribution $f(\sigma)$ (either continuous or discrete)
at a given packing fraction $\eta=\frac{\pi}{6}\rho \mth$, where
$\rho$ is the (total) number density and
\beq
\mn\equiv\langle
\sigma^n\rangle=\int_0^\infty d\sigma\, \sigma^n f(\sigma)
\label{1}
\eeq
is the $n$-th moment of the size distribution. Let
$g(\sigma,\sigma')$ denote the contact value of the pair correlation
function of particles of diameters $\sigma$ and $\sigma'$. This
function enters into the virial expression of the EOS as \cite{Lado}
\beqa
Z\equiv\frac{p}{\rho k_B
T}&=&1+4\frac{\eta}{\mth}\int_0^\infty d\sigma\int_0^\infty
d\sigma'\, f(\sigma)f(\sigma')
\left(\frac{\sigma+\sigma'}{2}\right)^3g(\sigma,\sigma')\nn
&=&1+\frac{\eta}{2
\mth}\left\langle\left(\sigma+\sigma'\right)^3g(\sigma,\sigma')\right\rangle,
\label{2}
\eeqa
where $Z$ is the compressibility factor, $p$ is the
pressure, $k_B$ is the Boltzmann constant and $T$ is the absolute
temperature. Assume further that the polydisperse hard-sphere
mixture may find itself in the presence of a hard wall. Since a hard
wall can be seen as a sphere of infinite diameter, the contact value
of the correlation function $g_w(\sigma)$ of a sphere of diameter
$\sigma$ with the wall is obtained from $g(\sigma,\sigma')$ as
\beq
g_w(\sigma)=\lim_{\sigma'\to\infty}g(\sigma,\sigma').
\label{3}
\eeq
Note that $g_w(\sigma)=\rho_w(\sigma)/\rho_{\text{bulk}}(\sigma)$
provides the ratio between the density of particles of size $\sigma$
adjacent to the wall, $\rho_w(\sigma)$, and the density of those
particles far away from the wall, $\rho_\text{bulk}(\sigma)$. There
is a sum rule connecting the pressure and the above contact values
\cite{Evans}, which provides an alternative route to the EOS, namely
\beq
Z_w=\int_0^\infty d\sigma\,f(\sigma)g_w(\sigma)=\langle
g_w(\sigma)\rangle,
\label{4}
\eeq
where the subscript $w$ in $Z_w$
has been used to emphasize that Eq.\ (\ref{4}) represents a route
alternative to the virial one, Eq.\ (\ref{2}), to get the EOS of the
hard-sphere polydisperse fluid. Our problem is then to compute
$g(\sigma,\sigma')$ and the associated $g_w(\sigma)$ for the
polydisperse hard-sphere mixture in the presence of a hard wall, so
that the condition $Z=Z_w$ is satisfied.

We consider a class of approximations of the type
\cite{Contact,SYH05}
\beq
g(\sigma,\sigma')=G(z(\sigma,\sigma')),
\label{n1} \eeq where \beq z(\sigma,\sigma')\equiv
\frac{2\sigma\sigma'}{\sigma+\sigma'}\frac{\mt}{\mth}
\label{7}
\eeq
is a dimensionless parameter. Therefore, at a given packing fraction
$\eta$, we are assuming that all the dependence of
$g(\sigma,\sigma')$ on $\sigma$, $\sigma'$ and on the details of the
size distribution $f(\sigma)$ is through the single parameter
$z(\sigma,\sigma')$. This implies that if two different pairs
$(\sigma_A,\sigma_A')$ and $(\sigma_B,\sigma_B')$ in two different
mixtures A and B (at the same packing fraction) have the same value
of the parameter $z$, \emph{i.e.},
$z_A(\sigma_A,\sigma_A')=z_B(\sigma_B,\sigma_B')$, then they also
have the same contact value of the rdf, \emph{i.e.},
$g_A(\sigma_A,\sigma_A')=g_B(\sigma_B,\sigma_B')$. The parameter
$z^{-1}(\sigma,\sigma')=(\sigma^{-1}+{\sigma'}^{-1})/(2\mu_2/\mu_3)$
can be interpreted as the arithmetic mean curvature, in appropriate
units, of spheres $\sigma$ and $\sigma'$ \cite{Contact}.

Notice that Eq.\ (\ref{n1}) implies in particular that
$g_w(\sigma)=G(z_w(\sigma))$, where $z_w(\sigma)=2 \sigma
{\mu_2}/{\mu_3}$. Once one accepts the ``universality'' ansatz
(\ref{n1}), the remaining problem lies in determining the form of
the function $G(z)$. This may be achieved by considering some
consistency conditions. Note that in the one-component limit,
\textit{i.e.}, $f(\sigma)=\delta(\sigma-\sigma_0)$, one has $z=1$,
so that \cite{SYH99,Contact}
\beq
G(z=1)=g_{\text{p}},
\label{n2}
\eeq
where $g_{\text{p}}$ is the contact value of the radial
distribution function of the one-component fluid at the same packing
fraction $\eta$ as the packing fraction of the mixture. Next, the
case of a mixture in which one of the species is made of point
particles, \textit{i.e.}, $\sigma \to 0$, leads to
\cite{SYH99,Contact,SYH05}
\beq
G(z=0)=\frac{1}{1-\eta}\equiv G_0.
\label{6}
\eeq
We now want consistency between both routes to the
EOS for any distribution $f(\sigma)$. To this end, we assume that
$z=0$ is a regular point, take into account condition (\ref{6}) and
expand $G(z)$ in a power series in $z$:
\beq
G(z)=G_0+\sum_{n=1}^\infty G_n z^n.
\label{n4}
\eeq
Using the ansatz (\ref{n1}) and Eq.\ (\ref{n4}) in Eq.\ (\ref{2})
one gets
\beqa
Z&=&1+\eta\sum_{n=0}^\infty 2^{n-1}G_n\frac{\mt^n}{\mth^{n+1}}
\langle\sigma^n{\sigma'}^n\left(\sigma+\sigma'\right)^{3-n}\rangle\nn
&=& G_0+3\eta\frac{\mo\mt}{\mth}G_0+\eta\sum_{n=1}^\infty
2^{n-1}G_n\frac{\mt^n}{\mth^{n+1}}
\langle\sigma^n{\sigma'}^n\left(\sigma+\sigma'\right)^{3-n}\rangle,
\label{n5}
\eeqa
where in the last step we have taken into account that
\beqa
1+\eta \frac{G_0}{2\mth}\langle
\left(\sigma+\sigma'\right)^{3}\rangle&=&1+\eta
\frac{G_0}{\mth}\left(\mth+3\mo\mt\right)\nn &=&G_0
+3\eta\frac{\mo\mt}{\mth}G_0.
\eeqa
Analogously, Eq.\ \eqref{4} becomes
\beq
Z_w=G_0+\sum_{n=1}^\infty
2^{n}G_n\frac{\mt^n}{\mth^{n}}\mu_n.
\label{n5bis}
\eeq
Notice that if the series \eqref{n4} is truncated after a given
order $n\geq 3$, $Z_w$ is given by the first $n$ moments of the size
distribution only. On the other hand, $Z$ still involves an infinite
number of moments if the truncation is made after $n\geq 4$ due to
the presence of terms like $\langle
\sigma^4{\sigma'}^4/(\sigma+\sigma')\rangle$ in Eq.\ \eqref{n5}.
Therefore, if we want the consistency condition $Z=Z_w$ to be
satisfied for \emph{any} polydisperse mixture, either the infinite
series \eqref{n4} needs to be considered or it must be truncated
after $n=3$. The latter is of course the simplest possibility and
thus we consider the approximation
\beq G(z)=G_0+G_1 z+G_2 z^2+G_3 z^3.
\label{5}
\eeq
As
a consequence, $Z$ and $Z_w$ depend functionally on $f(\sigma)$ only
through the first three moments (which is in the spirit of
Rosenfeld's Fundamental Measure Theory \cite{RosenfeldFMT}).

Using the approximation (\ref{5}) in Eqs.\ (\ref{n5}) and
(\ref{n5bis}) we are led to
\beq
Z=G_0+\eta\left[\frac{\mo
\mt}{\mth}
\left(3G_0+2G_1\right)+2\frac{\mt^3}{\mth^2}\left(G_1+2G_2+2G_3\right)\right],
\label{11} \eeq \beq Z_w=G_0+2\frac{\mo \mt}{\mth}G_1
+4\frac{\mt^3}{\mth^2}\left(G_2+2G_3\right).
\label{14}
\eeq

Thus far, the dependence of both $Z$ and $Z_w$ on the moments of
$f(\sigma)$ is explicit and we only lack the packing-fraction
dependence of $G_1$, $G_2$ and $G_3$. From Eqs.\ (\ref{11}) and
(\ref{14}) it follows that the difference between $Z$ and $Z_w$ is
given by
\beq
Z-Z_w=\frac{\mo \mt}{\mth}\left[3\eta
G_0-2(1-\eta)G_1\right] +2\frac{\mt^3}{\mth^2}\left[\eta
G_1-2(1-\eta)G_2-2(2-\eta)G_3\right].
\label{15}
\eeq
Therefore,
$Z=Z_w$ for any dispersity provided that
\beq
G_1=\frac{3 \eta}{2
\left(1-\eta\right)^2}, \label{n7a} \eeq \beq G_2=\frac{3 \eta^2}{4
\left(1-\eta\right)^3}-\frac{2-\eta}{1-\eta}G_3,
\label{n7b}
\eeq
where use has been made of the definition of $G_0$, Eq.\ (\ref{6}).
To close the problem, we use the equal size limit given in Eq.\
(\ref{n2}) and after a little algebra we are led to
\beq
G_2=(2-\eta)g_{\text{p}}-\frac{2+\eta^2/4}{\left(1-\eta\right)^2},
\label{n6a} \eeq \beq
G_3=(1-\eta)\left(g_{\text{p}}^{\text{SPT}}-g_{\text{p}}\right),
\label{n6b}
\eeq
where
\beq
g_{\text{p}}^{\text{SPT}}=\frac{1-\eta/2+\eta^2/4}{(1-\eta)^3}
\label{SPT}
\eeq
is the contact value of the radial distribution
function for a one-component fluid in the SPT. This completes our
derivation of the e3 approximation leading to the two following main
results for the contact values \cite{footnote1}:
\beqa
g(\sigma,\sigma')&=&\frac{1}{1-\eta}+\frac{3 \eta}{
\left(1-\eta\right)^2}\frac{\mt}{\mth}\frac{\sigma
\sigma'}{\sigma+\sigma'}+4\left[(2-\eta)g_{\text{p}}-\frac{2+\eta^2/4}{\left(1-\eta\right)^2}\right]\left(\frac{\mt}{\mth}\frac{\sigma
\sigma'}{\sigma+\sigma'}\right)^2\nn
&&+8(1-\eta)\left(g_{\text{p}}^{\text{SPT}}-g_{\text{p}}\right)\left(\frac{\mt}{\mth}\frac{\sigma
\sigma'}{\sigma+\sigma'}\right)^3,
\label{e3}
\eeqa
\beqa
g_w(\sigma)&=&\frac{1}{1-\eta}+\frac{3 \eta}{
\left(1-\eta\right)^2}\frac{\mt}{\mth}\sigma+4\left[(2-\eta)g_{\text{p}}-\frac{2+\eta^2/4}{\left(1-\eta\right)^2}\right]\left(\frac{\mt}{\mth}\sigma\right)^2\nn
&&+8(1-\eta)\left(g_{\text{p}}^{\text{SPT}}-g_{\text{p}}\right)\left(\frac{\mt}{\mth}\sigma\right)^3.
\label{e3w}
\eeqa
The label e3 is meant to indicate that  (i) the resulting contact
values are an \emph{extension} of the one-component contact value
$g_{\text{p}}$ and that (ii) $G(z)$ is a \emph{cubic} polynomial in
$z$. As mentioned earlier, all the theoretical proposals that also
comply with the universality ansatz (\ref{n1}), namely the PY, SPT,
BGHLL and our two former proposals \cite{SYH99,Contact} for the
contact values of the rdf , may be written in the form of Eq.\
(\ref{5}) with $G_3=0$, but only the SPT values also yield $Z=Z_w$
for any dispersity (see Table I in Ref.\ \cite{SYH05} for details).
Note further that the practical application of Eqs.\ (\ref{e3}) and
(\ref{e3w}) needs only the specification of the size distribution
$f(\sigma)$ and the choice of an approximate  expression for
$g_{\text{p}}$. For the latter, we will use the Carnahan--Starling
EOS \cite{CS69}, namely
\beq
g_{\text{p}}^{\text{CS}}=\frac{1-\eta/2}{(1-\eta)^3}
\label{CS}
\eeq
and use the notation eCS3 to label the approximation. As for the
size distribution, we will consider three cases:
\begin{enumerate}
\item
The top-hat distribution of sizes given by
\beq
f(\sigma)=\begin{cases} 1/2c,& \mo(1-c)\leq\sigma\leq\mo(1+c)\\
0,&\text{otherwise},
\end{cases}
\label{th}
\eeq
\item
The Schulz distribution of the form
 \beq
f(\sigma)=\frac{q+1}{q!\mo}\left(\frac{q+1}{\mo}\sigma\right)^q
\exp\left(-\frac{q+1}{\mo}\sigma\right)
\label{S}
\eeq
\item
The case of a bidisperse mixture, namely
\beq
f(\sigma)= x_1 \delta(\sigma-\sigma_1)+x_2
\delta(\sigma-\sigma_2),\quad x_1=1-x_2.
\label{D}
\eeq
\end{enumerate}
This choice of size distributions may seem to be to some extent
arbitrary (one could for instance have also included a log-normal
distribution). It has been mainly motivated by our desire to compare
with the (to our knowledge) available simulation data for
polydisperse hard-sphere mixtures in the presence of a hard planar
wall. Moreover, those simulations  have been computed for common
packing fractions in the polydisperse systems \eqref{th}--\eqref{D}.
In Table \ref{TableM} we present the values of the parameters
corresponding to the polydisperse mixtures that have been recently
studied, via Monte Carlo (MC) simulations by Buzzacchi \textit{et
al.} \cite{Ignacio} and the bidisperse mixtures studied also using
MC simulations by Malijevsk\'y \cite{Alexander}. We are now in a
position to assess the merits and limitations of our proposal.

\begin{table}
  \tbl{Parameters of the size distributions for the examined mixtures.}
{\begin{tabular}{@{}ccc} \toprule
Type&Parameters&$\eta$\\
\colrule
Top-hat &$c=0.2$&0.2\\
Top-hat &$c=0.2$&0.4\\
Top-hat &$c=0.7$&0.4\\
Schulz&$q=5$&0.2\\
Schulz&$q=5$&0.4\\
Bidisperse &$\sigma_2/\sigma_1=3, x_2=0.02$&0.206\\
Bidisperse &$\sigma_2/\sigma_1=3, x_2=0.04$&0.207\\
Bidisperse &$\sigma_2/\sigma_1=3, x_2=0.06$&0.208\\
Bidisperse &$\sigma_2/\sigma_1=3, x_2=0.0193$&0.404\\
Bidisperse &$\sigma_2/\sigma_1=3, x_2=0.0358$&0.407\\
Bidisperse &$\sigma_2/\sigma_1=3, x_2=0.0621$&0.401\\
\botrule
\end{tabular}}
\label{TableM}
\end{table}

\section{Comparison with simulation results}
\label{sec3}

\begin{figure}
\centerline{\epsfbox{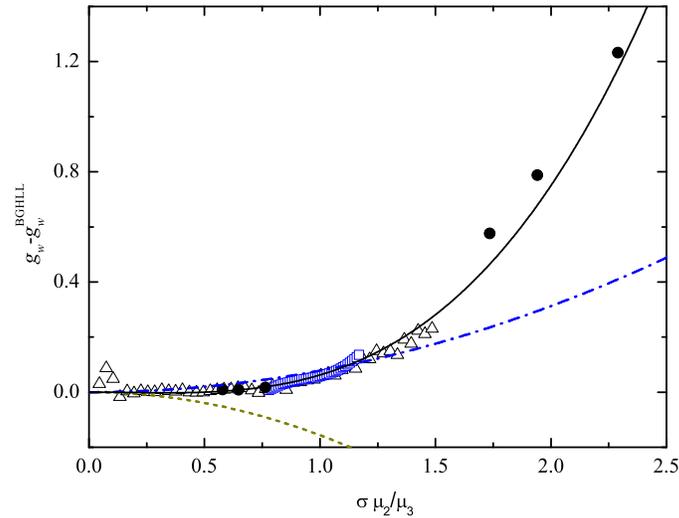}}
 \caption{Plot of the difference of
contact values $g_w-g_w^{\text{BGHLL}}$ as a function of
$\sigma{\mu_2}/{\mu_3}$ for the different polydisperse mixtures at
the fixed packing fraction $\eta=0.2$. The symbols are MC
simulations: top-hat distribution with $c=0.2$ (Ref.\
\protect\cite{Ignacio}) (squares), Schulz distribution with $q=5$
(Ref.\ \protect\cite{Ignacio}) (triangles), bidisperse mixtures
(Ref.\ \protect\cite{Alexander}) (filled circles). The lines are
PY({- - -}), SPT (--- $\cdot$ --- $\cdot$) and eCS3 (---).
\label{fig1}}
\end{figure}

\begin{figure}[htb]
 \centerline{\epsfbox{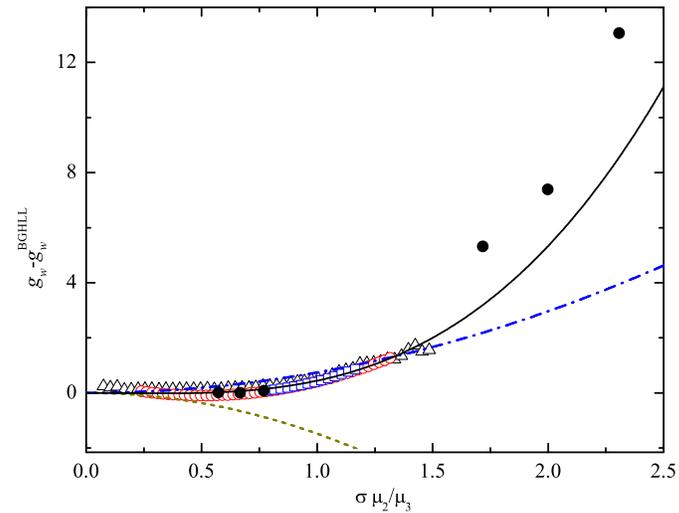}}
\caption{Plot of the difference of contact values
$g_w-g_w^{\text{BGHLL}}$ as a function of $\sigma{\mu_2}/{\mu_3}$
for the different polydisperse mixtures at the fixed packing
fraction $\eta=0.4$. The symbols are MC simulations: top-hat
distribution with $c=0.2$ (Ref.\ \protect\cite{Ignacio}) (squares),
top-hat distribution with $c=0.7$ (Ref.\ \protect\cite{Ignacio})
(circles), Schulz distribution with $q=5$ (Ref.\
\protect\cite{Ignacio}) (triangles), bidisperse mixtures (Ref.\
\protect\cite{Alexander}) (filled circles). The lines are PY({- -
-}), SPT (--- $\cdot$ --- $\cdot$) and eCS3 (---).
\label{fig2}}
\end{figure}

In Figs.\ \ref{fig1} and \ref{fig2} we display the comparison
between the results of various approximate theories for the contact
values of the wall-particle correlation functions and those obtained
from computer simulations for the mixtures given in Table
\ref{TableM} \cite{footnote2}. In view of the fact that our main
concern is to assess the universality ansatz, we have chosen to
represent the difference $g_w(\sigma)-g_w^\text{BGHLL}$ as a
function of $z_w(\sigma)/2=\sigma {\mu_2}/{\mu_3}$. The figures
suggest that, although the simulation data that we have examined are
only a few, the universality ansatz seems to be followed by them to
a large extent, thus providing an \emph{a posteriori} support to the
theoretical approaches that have this feature. In particular, the
data corresponding to the several polydisperse and bidisperse
mixtures with $\sigma \mt/\mth\lesssim 1.5$ overlap reasonably well.
In addition, the isolated points corresponding to the data for the
big spheres in the bidisperse mixtures are consistent with the trend
shown by the points with $\sigma \mt/\mth\lesssim 1.5$. Of course
more simulations for other values of $\sigma\mu_2/\mu_3$, especially
in the region $\sigma\mu_2/\mu_3\gtrsim 1.5$ or including ternary
systems with higher diameter ratios (which would offer a more
stringent test at higher values of $z$), would be welcome to further
confirm this assertion.

Regarding the theoretical approaches, as clearly seen in the figures
and already mentioned in Ref.\ \cite{SYH05}, the overall trend is
captured best by the eCS3 approach, while  the PY approximation
values are even qualitatively at odds with the simulation data. The
SPT overestimates the contact values in the region $\sigma \mt/\mth
\lesssim 1$ but it becomes the second best approximation for larger
values of $\sigma \mt/\mth$. In the latter region, all the theories
underestimate the simulation data, the eCS3 predictions being the
most accurate, especially for the smallest packing fraction.

\section{Concluding remarks}
\label{sec4}

In this paper we have examined the universality assumption that is
present in many theoretical derivations by which, once the packing
fraction is fixed, for all pairs of like and unlike spheres in a
polydisperse hard-sphere mixture with an arbitrary size distribution
and in the presence of a hard wall, the dependence of the contact
values of the particle-particle correlation function,
$g(\sigma,\sigma')$, and of the wall-particle correlation function,
$g_w(\sigma)$, on the diameters and on the composition is only
through a single dimensionless parameter and holds for an arbitrary
number of components. This was done by comparison with available
 MC simulation results for $g_w(\sigma)$ because, since $z_w(\sigma)>z(\sigma,\sigma')$, these contact
values represent a more stringent test for the universality ansatz
than the values of $g(\sigma,\sigma')$. While our analysis is
limited due to the few data that are at hand, the results suggest
that indeed the simulation data seem to comply reasonably well with
the ansatz, the results corresponding to the three different
bidisperse mixtures virtually falling on top of the polydisperse
ones for common values of $z_w$. The results also indicate that our
eCS3 approximation does a rather reasonable job. Although it
underestimates  the simulation data for high values of the parameter
$z_w$, it is still better than the other approximations sharing the
same universality property.

A noteworthy aspect of the comparison between the simulation data
and the theoretical approximations is that those proposals that
fulfill the condition $Z=Z_w$, namely the SPT and eCS3, are the ones
that show the best performance for high $z_w$. Since, as shown by
Eq.\ \eqref{n6b} and discussed in Ref.\ \cite{SYH05},  our e3
approach becomes identical to the SPT one when the choice
$g_\pure=g_\pure^{\text{SPT}}$ instead of
$g_\pure=g_\pure^{\text{CS}}$ is made, we can interpret it as a
versatile and flexible generalization of SPT. We are fully aware
that, apart from the consistency conditions that we have used, there
exist extra ones (see for instance Ref.\ \cite{HCorti04}) that one
might use as well within our approach. Assuming that the ansatz
(\ref{n1}) still holds, these conditions are related to the
derivatives of $G$ with respect to $z$, namely
\beq
\left.\frac{\partial G(z)}{\partial
z}\right|_{z=0}=\frac{3\eta}{2(1-\eta)^2}, \label{exC1} \eeq \beq
\left.\frac{\partial^2 G(z)}{\partial
z^2}\right|_{z=0}=\frac{3\eta}{1-\eta}\left(g_{\text{PY}}-\frac{1}{2}g\right),
\label{exC2} \eeq \beq \left.\frac{\partial^3 G(z)}{\partial
z^3}\right|_{z=2}=0,
\label{exC3}
\eeq
{where
$g_\pure^{\text{PY}}=(1+\eta/2)/(1-\eta)^2$ is the contact value of
the one-component hard-sphere fluid in the PY approximation. The
question immediately arises as to whether the fulfillment of these
extra conditions might influence the results we have presented in
this paper. Interestingly enough, as shown by Eq.\ \eqref{n7a},
condition (\ref{exC1}) is already satisfied by our e3 approximation
without having to be imposed. On the other hand, condition
(\ref{exC3}) implies $G_3=0$ in the e3 scheme and thus it is only
satisfied if $g_{\text{p}}=g_\pure^\text{SPT}$, in which case we
recover the SPT. Condition \eqref{exC2} is not fulfilled either by
the SPT or by the e3 approximation (except for a particular
expression of $g_\pure$ which is otherwise not very accurate). Thus,
fulfilling the extra conditions \eqref{exC2} and \eqref{exC3} with a
free $g_\pure$ requires either considering a higher order polynomial
in $z$ (in which case the consistency condition $Z=Z_w$ cannot be
satisfied for
 arbitrary mixtures, as discussed before) or not using the universality ansatz at all. In
the first case, we have checked that  a quartic or even a quintic
polynomial does not improve matters, whereas giving up the
universality assumption increases significantly the number of
parameters to be determined and seems not to be adequate in view of
the behavior observed in the simulation data. Therefore, e3 appears
to be a very reasonable compromise between simplicity and accuracy,
with the added bonus of being versatile to accommodate any choice
for $g_{\text{p}}$.

Finally, one should point out that the fact that the simulation
results give support to the validity of the universality assumption,
opens up the possibility of gaining information of rather
complicated polydisperse mixtures from the knowledge of simpler
systems using an approximation like our e3 approximation.

We want to thank  Dr.\ Alexandr Malijevsk\'y for making his
simulation results available to us prior to publication and Dr.\
David Reguera for bringing  Ref.\ \cite{HCorti04} to our attention.
 M.L.H. acknowledges the financial support of
DGAPA-UNAM under project IN-110406. The research of A.S. and S.B.Y.
has also been supported by the Ministerio de Educaci\'on y Ciencia
(Spain) through grant No.\ FIS2004-01399 (partially financed by
FEDER funds).


\begin{thebibliography}{99}
\bibitem{footnote1} Of course one may use these values to immediately write the
compressibility factor $Z=Z_w$ using either Eq.\ (\protect\ref{11})
or Eq.\ (\protect\ref{14}). The corresponding expression will be
ommitted since it is not required for the purposes of the present
paper but may be found in Ref.\ \protect\cite{SYH05}.
\bibitem{footnote2} Notice that the actual packing fractions of the
bidisperse mixtures quoted in Table \protect\ref{TableM} are rather
close but not exactly $\eta=0.2$ and $\eta=0.4$. One should expect
nevertheless that these minor differences should have little
influence on our analysis of the overall trends.
\end{thebibliography}

\end{document}